\documentclass[draftclsnofoot,onecolumn]{IEEEtran}
\usepackage{fullpage}
\usepackage{citesort}

\usepackage{epsfig}
\usepackage{graphicx}
\usepackage{caption}
\usepackage{subcaption}
\usepackage{fancybox}
\usepackage{color}
\usepackage{verbatim}
\usepackage{amssymb}

\usepackage[cmex10]{amsmath}
\usepackage{amsfonts}
\usepackage{cases}
\usepackage{url}

\usepackage{epstopdf}
\usepackage{amsthm}

\graphicspath{{./} {img/}}
\newtheorem{THEO}{Theorem}

\newtheorem*{REMA}{Remark}
\newtheorem{LEMM}{Lemma}

\def \xhatell {{\bf \widehat{x}}_{\ell_2}}

\begin{document}
%
\title{
\vspace*{-5mm}
Wiener Filters in Gaussian Mixture Signal Estimation with $\ell_\infty$-Norm Error}
%
%
%

\author{Jin~Tan,~\IEEEmembership{Student Member,~IEEE,}
Dror~Baron,~\IEEEmembership{Senior Member,~IEEE,}
~and~Liyi~Dai,~\IEEEmembership{Fellow,~IEEE}
\thanks{This work was supported in part by the National Science Foundation under Grant CCF-1217749 and in part by the U.S. Army Research Office under Grant W911NF-04-D-0003. Portions of the work appeared at the Information Theory and Applications workshop (ITA), San Diego, CA, Feb. 2013~\cite{TB2013ITA}, and at the IEEE Conference on Information Sciences and Systems (CISS), Princeton, NJ, Mar. 2014~\cite{Tan2014CISS}.
}
\thanks{Jin Tan and Dror Baron are with the Department of Electrical and Computer Engineering, North Carolina State University, Raleigh, NC 27695. E-mail: \{jtan, barondror\}@ncsu.edu. Liyi Dai is with the Computing Sciences Division, US Army Research Office, Research Triangle Park, NC 27709. E-mail:liyi.dai.civ@mail.mil.}
}
\maketitle \thispagestyle{empty}
\newcommand{\xhat}{\widehat{\mathbf{x}}}
\newcommand{\xhati}{\widehat{x}_i}
\def\x{{\mathbf x}}
\def\L{{\cal L}}
\begin{abstract}
Consider the estimation of a signal~${\bf x}\in\mathbb{R}^N$ from noisy observations~${\bf r=x+z}$, where 
the input~${\bf x}$ is generated by an independent and identically distributed (i.i.d.) Gaussian mixture source, and ${\bf z}$ is additive white Gaussian noise (AWGN) in parallel Gaussian channels. Typically, the~$\ell_2$-norm error (squared error) is used to quantify the performance of the estimation process. In contrast, we consider the~$\ell_\infty$-norm error (worst case error). For this error metric, we prove that, in an asymptotic setting where the signal dimension~$N\to\infty$, the~$\ell_\infty$-norm error always comes from the Gaussian component that has the largest variance, and the Wiener filter asymptotically achieves the optimal expected~$\ell_\infty$-norm error. The i.i.d. Gaussian mixture case is easily applicable to i.i.d. Bernoulli-Gaussian distributions, which are often used to model sparse signals. Finally, our results can be extended to linear mixing systems with i.i.d. Gaussian mixture inputs, 
in settings where a linear mixing system can be decoupled to parallel Gaussian channels.
\end{abstract}


%
\IEEEpeerreviewmaketitle

\section{Introduction}
\label{sec:intro}
\subsection{Motivation}
\label{subsec:motivation}
The Gaussian distribution is widely used to describe the probability densities of various types of data, owing to its advantageous mathematical properties~\cite{Papoulis91}. It has been shown that non-Gaussian distributions can often be sufficiently approximated by an infinite mixture of Gaussians~\cite{Alecu2006}, so that the mathematical advantages of the Gaussian distribution can be leveraged when discussing non-Gaussian signals~\cite{Bijaoui2002,Alecu2006,Tabuchi2007,EMGMCISS,EMGMTSP}. 
In practice, signals are often contaminated by noise during sampling or transmission, and therefore estimation of signals from their noisy observations are needed.
Most estimation methods evaluate the performance by the ubiquitous~$\ell_2$-norm error~\cite{Alecu2006} (squared error).
However, there are applications where other error metrics may be preferred~\cite{Tan2014}. 
For example, the~$\ell_2$ error criterion ensures that the estimated signal has low square error on average, but does not guarantee that every estimated signal component is close to the corresponding original signal component. In problems such as image and video compression~\cite{Dalai2006} where the reconstruction quality at every signal component is important, it would be better to optimize for~$\ell_\infty$-norm error.
Our interest in the $\ell_{\infty}$-norm error is also motivated by applications including
wireless communications~\cite{Studer2012},
group testing~\cite{Gilbert2012} and trajectory planning in control systems~\cite{Egerstedt1999},
where we want to decrease the worst-case sensitivity to noise.

\subsection{Problem setting}
\label{subsec:Prob}

In this correspondence, our main focus is on {\em parallel Gaussian channels}, and the results can be extended to {\em linear mixing systems}. In both settings, the input~$\x\in\mathbb{R}^N$ is generated by an independent and identically distributed (i.i.d.) {\em Gaussian mixture} source,
\begin{equation}
x_i\sim \sum_{k=1}^Ks_k\cdot\mathcal{N}(\mu_k,\sigma_k^2)
=\sum_{k=1}^K
\frac{s_k}{\sqrt{2\pi\sigma_k^2}}e^{-\frac{(x_i-\mu_k)^2}{2\sigma_k^2}},
\label{eq:MGauss}
\end{equation}
where
the subscript~$(\cdot)_i$ denotes the $i$-th component of a sequence (or a vector),~$\mu_1,\mu_2,\ldots,\mu_K$ (respectively, $\sigma_1^2$, $\sigma_2^2$, $\ldots$,$\sigma_K^2$) are the means (respectively, variances) of the Gaussian components, and~$0<s_1,s_2,\ldots,s_K<1$ are the probabilities of the~$K$ Gaussian components. Note that~$\sum_{k=1}^Ks_k=1$.  A special case of the Gaussian mixture is Bernoulli-Gaussian,
\begin{equation}
x_i\sim s\cdot\mathcal{N}(\mu_x,\sigma_x^2)+(1-s)\cdot\delta(x_i),
\label{eq:BGauss}
\end{equation}
for some~$0<s<1$,~$\mu_x$, and~$\sigma_x^2$, where~$\delta(\cdot)$ is the delta function~\cite{Papoulis91}. The zero-mean Bernoulli-Gaussian model is often used in sparse signal processing~\cite{CandesRUP,DonohoCS,GuoWang2007,GuoWang2008,Rangan2010CISS,Rangan2010,Starck2010,VilaSchniter2011}.

In {\em parallel Gaussian channels}~\cite{Tabuchi2007,Bijaoui2002}, we consider
\begin{equation}
{\bf r=x+z},
\label{eq:scalar}
\end{equation}
where~${\bf r,x,z}\in\mathbb{R}^N$ are the output signal, the input signal, and the additive white Gaussian noise (AWGN), respectively. The AWGN channel can be described by the conditional distribution
\begin{equation}
f_{\bf R|X}({\bf r|x})=\prod_{i=1}^Nf_{R|X}(r_i|x_i)=\prod_{i=1}^N\frac{1}{\sqrt{2\pi\sigma_z^2}}\exp{\left(-\frac{(r_i-x_i)^2}{2\sigma_z^2}\right)},
\label{eq:Gchannel}
\end{equation}
where~$\sigma_z^2$ is the variance of the Gaussian noise.

In a {\em linear mixing system}~\cite{CandesRUP,DonohoCS,GuoWang2008,Rangan2010CISS}, we consider
\begin{equation}
\label{eq:basicSystem}
\mathbf{w=\Phi x},
\end{equation}
the {\em measurement matrix}~$\mathbf{\Phi}\in\mathbb{R}^{M\times N}$ is sparse and its entries are i.i.d. Because each component of the measurement vector~$\mathbf{w}\in\mathbb{R}^M$ is a linear combination of the components of~$\mathbf{x}$, we call the system~\eqref{eq:basicSystem} a {\em linear mixing system}. The measurements~$\mathbf{w}$ are passed through a bank of separable scalar channels characterized by conditional distributions
\begin{eqnarray}
f_{\mathbf{Y|W}}(\mathbf{y|w})=\prod_{i=1}^M f_{Y|W}(y_i|w_i),
\label{eq:DisChannel}
\end{eqnarray}
where~$\mathbf{y}\in\mathbb{R}^M$ are the channel outputs. However, unlike the parallel Gaussian channels~\eqref{eq:Gchannel}, the channels~\eqref{eq:DisChannel} of the linear mixing system are not restricted to Gaussian \cite{GuoWang2007,Rangan2010CISS,Rangan2010}. 

Our goal is to estimate the original input signal~$\mathbf{x}$ either from the parallel Gaussian channel outputs~${\bf r}$ in~\eqref{eq:scalar} or from the linear mixing system outputs~$\mathbf{y}$ and the measurement matrix $\mathbf{\Phi}$ 
in~\eqref{eq:basicSystem} and~\eqref{eq:DisChannel}. 
To evaluate how accurate the estimation process is, 
we quantify the~$\ell_\infty$-norm error between $\mathbf{x}$ and its estimate~$\xhat$,
\begin{equation*}
\|\widehat{\mathbf{x}}-\mathbf{x}\|_{\infty} = 
\max_{i \in \{1,\ldots,N\} } |\widehat{x}_i - x_i|;
\end{equation*}
this error metric helps prevent any significant errors during the estimation process. The estimator that minimizes the expected value of~$\|{\bf\widehat{x}-x}\|_\infty$ is called the {\em minimum mean~$\ell_\infty$-norm error estimator}. We denote this estimator by~${\bf\widehat{x}_{\ell_\infty}}$, which can be expressed as
\begin{equation}
{\bf\widehat{x}}_{\ell_\infty}=\arg\min_{\bf\widehat{x}}E\left[\|{\bf\widehat{x}-x}\|_\infty\right].
\label{eq:ell_infty_estimator}
\end{equation}

\subsection{Related work}
\label{subsec:relatedWork}

Gaussian mixtures are widely used to model various types of signals, and a number of signal estimation methods have been introduced to take advantage of the Gaussian mixture distribution. For example,
an infinite Gaussian mixture model was proposed in~\cite{Alecu2006} to represent real data such as images, and a denoising scheme based on local linear estimators was developed to estimate the original data. A similar algorithm based on an adaptive Wiener filter was applied to denoise X-ray CT images~\cite{Tabuchi2007}, where a Gaussian mixture model was utilized. However, these works only quantified the $\ell_2$-norm error of the denoising process.
Signal estimation problems with~$\ell_\infty$-norm error have not been well-explored, but there have been studies on general properties of the $\ell_\infty$-norm. For example, in Clark~\cite{clark1961}, the author developed a deductive method to calculate the distribution of the greatest element in a finite set of random variables; and Indyk~\cite{indyk2001} discussed how to find the nearest neighbor of a point while taking the $\ell_\infty$-norm distance into consideration. 

\subsection{Contributions}
\label{sec:contrib}

In this correspondence, we study the estimator that minimizes the~$\ell_\infty$-norm error in parallel Gaussian channels
in an asymptotic setting where the signal dimension $N \to \infty$.
We prove that, when estimating an input signal that is generated by an i.i.d. Gaussian mixture source, the~$\ell_\infty$-norm error always comes from the Gaussian component that has the largest variance. Therefore, the well-known Wiener filter achieves the minimum mean~$\ell_\infty$-norm error. The Wiener filter is a simple linear function that is applied to the channel outputs, where the multiplicative constant of the linear function is computed by considering the greatest variance of the Gaussian mixture components~\eqref{eq:MGauss} and the variance of the channel noise. Moreover, the Wiener filter can be applied to linear mixing systems defined in~\eqref{eq:basicSystem} and~\eqref{eq:DisChannel} to minimize the~$\ell_\infty$-norm error, based on settings where a linear mixing system can be decoupled to parallel Gaussian channels~\cite{GuoVerdu2005,Montanari2006,GuoBaronShamai2009,RFG2012,DMM2009,Rangan2010CISS,Rangan2010}.

The remainder of the correspondence is arranged as follows. We provide our main results and discuss their applications in Section \ref{sec:main}. Proofs of the main results appear in Section~\ref{sec:proofs}, while Section~\ref{sec:concld} concludes.

\section{Main Results}
\label{sec:main}

For parallel Gaussian channels~\eqref{eq:scalar}, the {\em minimum mean squared error estimator}, denoted by~$\xhatell$, is achieved by the conditional expectation~$E[{\bf x|r}]$. 
If the input signal~${\bf x}$ is i.i.d. Gaussian (not a Gaussian mixture), i.e.,~${x_i}\sim\mathcal{N}(\mu_x,\sigma_x^2)$, then the estimate
\begin{equation}
\xhatell=E[{\bf x|r}]=\frac{\sigma_x^2}{\sigma_x^2+\sigma_z^2}({\bf r}-\mu_x)+\mu_x
\label{eq:WF}
\end{equation}
achieves the minimum mean squared error, where~$\sigma_z^2$ is the variance of the Gaussian noise~${\bf z}$ in~\eqref{eq:scalar}, and we use the convention that adding a scalar to (respectively, subtracting a scalar from) a vector means adding this scalar to (respectively, subtracting this scalar from) every component of the vector.
This format in~\eqref{eq:WF} is called the {\em Wiener filter} in signal processing~\cite{Wiener1949}.
It has been shown by Sherman~\cite{Sherman1955,Sherman1958} that, besides the~$\ell_2$-norm error, the linear Wiener filter is also optimal for all~$\ell_p$-norm errors ($p\ge1$), including the~$\ell_\infty$-norm error. Surprisingly, we find that, if the input signal is generated by an i.i.d. Gaussian mixture source, then the Wiener filter asymptotically minimizes the expected~$\ell_\infty$-norm error.

Before providing the result for the Gaussian mixture input case, which is mathematically involved, we begin with an analysis of the simpler Bernoulli-Gaussian input case.

\begin{THEO}
\label{thm:01}
In parallel Gaussian channels~\eqref{eq:scalar}, if the input signal~${\bf x}$ is generated by an i.i.d. Bernoulli-Gaussian source defined in~\eqref{eq:BGauss}, then the Wiener filter 
\begin{equation}
{\bf \widehat{x}}_{\text{W,BG}}=\frac{\sigma_x^2}{\sigma_x^2+\sigma_z^2}({\bf r}-\mu_x)+\mu_x
\label{eq:Wiener01}
\end{equation}
asymptotically achieves the minimum mean~$\ell_\infty$-norm error. More specifically,
\begin{equation}
\lim_{N\to\infty}\frac{E\left[\|\mathbf{x}-\widehat{\mathbf{x}}_\text{W,BG}\|_\infty
\right]}
{E\left[\|\mathbf{x}-\widehat{\mathbf{x}}_{\ell_\infty}\|_\infty
\right]}
=1,\nonumber
\end{equation}
where~$\widehat{\bf x}_{\ell_\infty}$ satisfies~\eqref{eq:ell_infty_estimator}.
\end{THEO}

Theorem~\ref{thm:01} is proved in Section~\ref{appen:ThmProof}. 
The proof combines concepts in typical sets~\cite{Cover06} and a result by Gnedenko~\cite{Gnedenko1943}, which provided asymptotic properties of the maximum of a Gaussian sequence.
The main idea of the proof is to show that with overwhelming probability the maximum absolute error satisfies~$\|{\bf x-\widehat{x}}\|_\infty=|x_i-\widehat{x}_i|$, where $i\in\mathcal{I}=\{i:x_i\sim\mathcal{N}(\mu_x, \sigma_x^2)\}$, 
i.e., $\mathcal{I}$ is the index set that includes all the Gaussian components of the vector~${\bf x}$, and excludes all the zero components of~${\bf x}$.
Therefore, minimizing~$\|{\bf x-\widehat{x}}\|_\infty$ is equivalent to minimizing~$\|{\bf x_\mathcal{I}-\widehat{x}_\mathcal{I}}\|_\infty$, where~$(\cdot)_\mathcal{I}$ denotes a subvector with entries in the index set~$\mathcal{I}$. Because the vector ${\bf {x}_\mathcal{I}}$ is i.i.d. Gaussian, the Wiener filter minimizes~$\|{\bf x_\mathcal{I}-\widehat{x}_\mathcal{I}}\|_\infty$~\cite{Sherman1955,Sherman1958}; hence the Wiener filter minimizes~$\|{\bf x-\widehat{x}}\|_\infty$ with overwhelming probability. On the other hand, the cases where~$\|{\bf x-\widehat{x}}\|_\infty=|x_i-\widehat{x}_i|$ and $i\notin\mathcal{I}$ are rare, the mean~$\ell_\infty$-norm error of the Wiener filter barely increases, and so the Wiener filter asymptotically minimizes the expected~$\ell_\infty$-norm error.

Having discussed the Bernoulli-Gaussian case, let us proceed to the Gaussian mixture case defined in~\eqref{eq:MGauss}. Here the maximum absolute error between~${\bf x}$ and the estimate~${\bf \widehat{x}}$ satisfies~$\|{\bf x-\widehat{x}}\|_\infty=|x_i-\widehat{x}_i|$, where~$i\in\mathcal{I}'=\{i:x_i\sim\mathcal{N}(\mu_m,\sigma_m^2)\}$, and~$m=\arg\max_{k\in\{1,2,\ldots,K\}}\sigma_k^2$. That is, the maximum absolute error between~${\bf x}$ and~${\bf\widehat{x}}$ lies in an index that {\em corresponds to the Gaussian mixture component with greatest variance}.
\begin{THEO}
\label{coro:01}
In parallel Gaussian channels~\eqref{eq:scalar}, if the input signal~${\bf x}$ is generated by an i.i.d. Gaussian mixture source defined in~\eqref{eq:MGauss}, then the Wiener filter 
\begin{equation}
{\bf \widehat{x}}_{\text{W,GM}}=\frac{\sigma_m^2}{\sigma_m^2+\sigma_z^2}\left({\bf r}-\mu_m\right)+\mu_m
\label{eq:Wiener02}
\end{equation}
asymptotically achieves the minimum mean~$\ell_\infty$-norm error, where~$m=\arg\max_{k\in\{1,2,\ldots,K\}}\sigma_k^2$. More specifically,
\begin{equation}
\lim_{N\to\infty}\frac{E\left[\|\mathbf{x}-\widehat{\mathbf{x}}_\text{W,GM}\|_\infty
\right]}
{E\left[\|\mathbf{x}-\widehat{\mathbf{x}}_{\ell_\infty}\|_\infty
\right]}
=1,\nonumber
\end{equation}
where~$\widehat{\bf x}_{\ell_\infty}$ satisfies~\eqref{eq:ell_infty_estimator}.
\end{THEO}

The proof of Theorem~\ref{coro:01} is given in Section~\ref{appen:coro_01}.
We note in passing that the statements in Theorems~\ref{thm:01} and~\ref{coro:01} do not hold for~$\ell_p$-norm error ($0<p<\infty$). Because there is a one to one correspondence between the parameters ($\mu_k$ and $\sigma_k^2$) of a Gaussian mixture component and its corresponding Wiener filter, if a Wiener filter is optimal in the $\ell_p$ error sense for any of the Gaussian mixture components, then it is suboptimal for the rest of the mixture components. Therefore, any single Wiener filter is suboptimal in the $\ell_p$ error sense for any Gaussian mixture signal comprising more than one Gaussian component.

\begin{REMA}
Theorems~\ref{thm:01} and~\ref{coro:01} can be extended to linear mixing systems. We consider a linear mixing system defined in~\eqref{eq:basicSystem}, where the matrix~${\bf\Phi}\in\mathbb{R}^{M\times N}$ is i.i.d. sparse, and let $\Gamma$ denote the average number of nonzeros in each row of~${\bf\Phi}$. It has been shown~\cite{GuoVerdu2005,Montanari2006,GuoWang2007,GuoBaronShamai2009,RFG2012,DMM2009,Rangan2010CISS,Rangan2010} that,
in a large-sparse-limit where $M, N, \Gamma\to\infty$ with $M/N\to\beta<\infty$ for some constant $\beta$ and $\Gamma=o(N^{1/2})$, a linear mixing system~\eqref{eq:basicSystem} and~\eqref{eq:DisChannel} can be decoupled to an equivalent set of parallel Gaussian channels,
$
{\bf q = x + v}
$
where~${\bf v}\in\mathbb{R}^N$ is the equivalent Gaussian noise, and~${\bf q}\in\mathbb{R}^N$ are the outputs of the decoupled parallel Gaussian channels. 
The statistical properties of the noise~${\bf v}$ are characterized by Tanaka's fixed point equation~\cite{Tanaka2002,GuoVerdu2005,Montanari2006,GuoBaronShamai2009}. Therefore, when the input signal~${\bf x}$ is generated by an i.i.d. Gaussian mixture source, by applying the Wiener filter to~${\bf q}$, we can obtain the estimate that minimizes the~$\ell_\infty$-norm error of the signal estimation process.
\end{REMA}

\section{Proofs}
\label{sec:proofs}

\subsection{Proof of Theorem~\ref{thm:01}}
\label{appen:ThmProof}

{\bf Two error patterns:} We begin by defining two error patterns. Consider parallel Gaussian channels~\eqref{eq:scalar}, where the input signal~$x_i\sim s\cdot\mathcal{N}(\mu_x,\sigma_x^2)+ (1-s)\cdot\delta(x_i)$ for some~$s$, and the noise~$z_i\sim\mathcal{N}(0,\sigma_z^2)$. The Wiener filter (linear estimator) for the Bernoulli-Gaussian input is~$\mathbf{\widehat{x}}_\text{W,BG}=\frac{\sigma_x^2}{\sigma_x^2+\sigma_z^2}\cdot (\mathbf{r}-\mu_x)+\mu_x$. Let $\mathcal{I}$ denote the index set where $x_i\sim\mathcal{N}(\mu_x,\sigma_x^2)$, and let $\mathcal{J}$ denote the index set where~$x_j\sim\delta(x_j)$. We define two types of error patterns: 
({\em i}) for
\begin{equation} i\in\mathcal{I}\triangleq\left\{i:x_i\sim\mathcal{N}\left(\mu_x,\sigma_x^2\right)\right\},\nonumber\\
\end{equation}
the error is 
\begin{equation}
e_i\triangleq\widehat{x}_{\text{W,BG},i}-x_i=\frac{\sigma_x^2}{\sigma_x^2+\sigma_z^2}\cdot (r_i-\mu_x)+\mu_x-x_i
\sim\mathcal{N}\left(0,\frac{\sigma_x^2\sigma_z^2 }{\sigma_x^2+\sigma_z^2}\right),
\nonumber
\end{equation} 
where we remind readers that~$\widehat{x}_{\text{W,BG},i}$ denotes the~$i$-th component of the vector~${\bf\widehat{x}_{\text{W,BG}}}$ in~\eqref{eq:Wiener01};
and ({\em ii}) for
\begin{equation}
j\in\mathcal{J}\triangleq\left\{j:x_j\sim\delta(x_j)\right\},
\nonumber
\end{equation}
the error is 
\begin{equation}
\widetilde{e}_j\triangleq\widehat{x}_{\text{W,BG},j}-x_j=\frac{\sigma_x^2}{\sigma_x^2+\sigma_z^2}\cdot (r_i-\mu_x)+\mu_x-0
\sim\mathcal{N}\left(\frac{\sigma_z^2}{\sigma_x^2+\sigma_z^2}\mu_x,
\frac{\sigma_x^4\sigma_z^2}{(\sigma_x^2+\sigma_z^2)^2}
\right).\nonumber
\end{equation} 

{\bf Maximum of error patterns: } Let us compare~$\max_{i\in\mathcal{I}} |e_i|$ and~$\max_{j\in\mathcal{J}}|\widetilde{e}_j|$.

\begin{LEMM}
\label{lemma}
Suppose~$u_i$ is an i.i.d. Gaussian sequence of length~$N$,~$u_i\sim\mathcal{N}(\mu,\sigma^2)$ for~$i\in\{1,2,\ldots,N\}$, then~$\frac{\max_{1\le i\le N}|u_i|}{\sqrt{2\sigma^2\cdot\ln (N)}}$ converges to 1 in probability. That is,
\begin{equation}
\lim_{N\rightarrow\infty}\Pr\left(\left|\frac{\max_{1\le i\le N}|u_i|}{\sqrt{2\sigma^2\cdot\ln (N)}}-1\right|<\Delta\right)=1,
\label{eq:maxE}
\end{equation}
for any~$\Delta>0$.
\end{LEMM}
Lemma~\ref{lemma} is proved in Section~\ref{append:lemma}.

Before applying Lemma~\ref{lemma}, we define a set~$A_\epsilon$ of~possible inputs ${\bf x}$ such that the numbers of components in the sets~$\mathcal{I}$ and~$\mathcal{J}$ both go to infinity as~$N\to\infty$,
\begin{equation}
A_\epsilon\triangleq\left\{{\bf x}:\left|\frac{|\mathcal{I}|}{N}-s\right|<\epsilon\right\},
\label{eq:A_eps}
\end{equation}
where~$\epsilon>0$ and~$\epsilon\to0$ (namely,~$\epsilon\to0^+$) as a function of signal dimension~$N$, and~$|\mathcal{I}|$ denotes the cardinality of the set~$\mathcal{I}$. The definition of~$A_\epsilon$ suggests that~$\left|\frac{|\mathcal{J}|}{N}-(1-s)\right|<\epsilon$ and~$|\mathcal{I}|+|\mathcal{J}|=N$. Therefore, if~${\bf x}\in A_\epsilon$, then~$|\mathcal{I}|,|\mathcal{J}|\to\infty$ as~$N\to\infty$.

Now we are ready to evaluate~$\max_{i\in\mathcal{I}} |e_i|$ and~$\max_{j\in\mathcal{J}}|\widetilde{e}_j|$. For i.i.d. Gaussian random variables $e_i\sim\mathcal{N}(0,\frac{\sigma_x^2\sigma_z^2}{\sigma_x^2+\sigma_z^2})$, where~$i\in\mathcal{I}$, the equality~\eqref{eq:maxE} in Lemma~\ref{lemma} becomes
\begin{equation}
\lim_{N\rightarrow\infty}\Pr\left(\left.\left|\frac{\max_{i\in\mathcal{I}}|e_i|}{\sqrt{2\cdot\frac{\sigma_x^2\sigma_z^2}{\sigma_x^2+\sigma_z^2}\cdot\ln (|\mathcal{I}|)}}-1\right|<\Delta\right|{\bf x}\in A_\epsilon\right)=1,\label{eq:maxE01}
\end{equation}
for any~$\Delta>0$. 
For i.i.d. Gaussian random variables $\widetilde{e}_j$, where~$j\in\mathcal{J}$, the equality~\eqref{eq:maxE} becomes
\begin{equation}
\lim_{N\rightarrow\infty}\Pr\left(\left.\left|\frac{\max_{j\in\mathcal{J}}\left|\widetilde{e}_j\right|}{\sqrt{2\cdot\frac{\sigma_x^4\sigma_z^2}{(\sigma_x^2+\sigma_z^2)^2}\cdot\ln (|\mathcal{J}|)}}-1\right|<\Delta\right|{\bf x}\in A_\epsilon\right)=1,\label{eq:maxE02}
\end{equation}
for any~$\Delta>0$.

Equations~\eqref{eq:maxE01} and~\eqref{eq:maxE02} suggest that
\begin{equation}
\lim_{N\to\infty} E\left[\left.\frac{\max_{i\in\mathcal{I}}|e_i|}{\sqrt{2
\cdot\frac{\sigma_x^2\sigma_z^2}{\sigma_x^2+\sigma_z^2}\cdot
\ln(|\mathcal{I}|)}}\right|{\bf x}\in A_\epsilon\right]=1\nonumber
\end{equation}
and
\begin{equation}
\lim_{N\to\infty} E\left[\left.\frac{\max_{j\in\mathcal{J}}|\widetilde{e}_j|}{\sqrt{2
\cdot\frac{\sigma_x^4\sigma_z^2}{(\sigma_x^2+\sigma_z^2)^2}\cdot
\ln(|\mathcal{J}|)}}\right|{\bf x}\in A_\epsilon\right]=1,\nonumber
\end{equation}
which yield
\begin{equation}
\lim_{N\to\infty} 
E\left[\left.\frac{\max_{i\in\mathcal{I}}|e_i|}{\sqrt{
\ln(N)}}\right|{\bf x}\in A_\epsilon\right]=
\lim_{N\to\infty} \sqrt{2
\cdot\frac{\sigma_x^2\sigma_z^2}{\sigma_x^2+\sigma_z^2}\cdot
\frac{\ln(|\mathcal{I}|)}{\ln(N)}}
\label{eq:Bounded_1}
\end{equation}
and
\begin{equation}
\lim_{N\to\infty} 
E\left[\left.\frac{\max_{j\in\mathcal{J}}|\widetilde{e}_j|}{\sqrt{
\ln(N)}}\right|{\bf x}\in A_\epsilon\right]=
\lim_{N\to\infty} \sqrt{2
\cdot\frac{\sigma_x^4\sigma_z^2}{(\sigma_x^2+\sigma_z^2)^2}\cdot
\frac{\ln(|\mathcal{J}|)}{\ln(N)}}.
\label{eq:Bounded_2}
\end{equation}
According to the definition of~$A_\epsilon$ in~\eqref{eq:A_eps}, where $s$ is a constant, and~$\epsilon\to0^+$,  
\begin{equation}
N(s-\epsilon)<|\mathcal{I}|<N(s+\epsilon) \quad\text{and} \quad N(1-s-\epsilon)<|\mathcal{J}|<N(1-s+\epsilon),
\label{eq:IJ_ineq}
\end{equation}
and thus
\begin{equation}
\lim_{N\to\infty}\sqrt{\frac{\ln{(|\mathcal{I}|)}}{\ln(N)}}
=1 \quad\text{and}\quad \lim_{N\to\infty}\sqrt{\frac{\ln{(|\mathcal{J}|)}}{\ln(N)}}
=1.\label{eq:ratio}
\end{equation}
Finally, equations~\eqref{eq:Bounded_1} and~\eqref{eq:Bounded_2} become
\begin{equation}
\lim_{N\to\infty} 
E\left[\left.\frac{\max_{i\in\mathcal{I}}|e_i|}{\sqrt{
\ln(N)}}\right|{\bf x}\in A_\epsilon\right]=
\sqrt{2
\cdot\frac{\sigma_x^2\sigma_z^2}{\sigma_x^2+\sigma_z^2}}
\label{eq:Bounded_exp_1}
\end{equation}
and
\begin{equation}
\lim_{N\to\infty} E\left[\left.\frac{\max_{j\in\mathcal{J}}|\widetilde{e}_j|}{\sqrt{
\ln(N)}}\right|{\bf x}\in A_\epsilon\right]=
\sqrt{2
\cdot\frac{\sigma_x^4\sigma_z^2}{(\sigma_x^2+\sigma_z^2)^2}}.
\label{eq:Bounded_exp_2}
\end{equation}
Combining~\eqref{eq:maxE01} and~\eqref{eq:maxE02},
\begin{equation}
\lim_{N\to\infty} \Pr\left(\left. 
\frac{1-\Delta}{1+\Delta}<\frac{\max_{i\in\mathcal{I}} |e_i|}
{\max_{j\in\mathcal{J}} |\widetilde{e}_j|}\cdot
\frac{\sqrt{2\cdot\frac{\sigma_x^4\sigma_z^2}{(\sigma_x^2+\sigma_z^2)^2}\cdot\ln(|\mathcal{J}|)}}
{\sqrt{2\cdot\frac{\sigma_x^2\sigma_z^2}{\sigma_x^2+\sigma_z^2}\cdot\ln(|\mathcal{I}|)}}<\frac{1+\Delta}{1-\Delta}
\right|{\bf x}\in A_\epsilon\right)
=1.
\label{eq:Range_ratio}
\end{equation}
Note that
\begin{equation}
\sqrt{\frac{\ln(N)+\ln(1-s-\epsilon)}{\ln(N)+\ln(s+\epsilon)}}=\sqrt{\frac{\ln(N(1-s-\epsilon))}{\ln(N(s+\epsilon))}}
<
\sqrt{\frac{\ln(|\mathcal{J}|)}{\ln(|\mathcal{I}|)}}
<
\sqrt{\frac{\ln(N(1-s+\epsilon))}{\ln(N(s-\epsilon))}}=
\sqrt{\frac{\ln(N)+\ln(1-s+\epsilon)}{\ln(N)+\ln(s-\epsilon)}}.\nonumber
\end{equation}
Then the following limit holds,
\begin{equation}
\lim_{N\to\infty}\sqrt{\frac{\ln (|\mathcal{J}|)}{\ln (|\mathcal{I}|)}}=1.\nonumber
\end{equation}
We can write the above limit in probabilistic form,
\begin{equation}
\lim_{N\to\infty}\Pr\left(\left.\left|\sqrt{\frac{\ln (|\mathcal{J}|)}{\ln (|\mathcal{I}|)}}-1\right|<\Delta\right|{\bf x}\in A_\epsilon\right)=1,\label{eq:J_over_I}
\end{equation}
for any $\Delta>0$.
Because of the logarithms in~\eqref{eq:J_over_I}, the ratio~$\frac{\sqrt{2\cdot\ln (|\mathcal{J}|)}}{\sqrt{2\cdot\ln (|\mathcal{I}|)}}$ is sufficiently close to~$1$ as~$N$ is astronomically large. This is why we point out in Section~\ref{sec:concld} that the asymptotic results in this correspondence might be impractical.
Plugging~\eqref{eq:J_over_I} into~\eqref{eq:Range_ratio},

\begin{equation}
\lim_{N\to\infty} \Pr\left(\left. 
\frac{1-\Delta}{(1+\Delta)^2}\cdot
\sqrt{\frac{\sigma_x^2+\sigma_z^2}{\sigma_x^2}}
<\frac{\max_{i\in\mathcal{I}} |e_i|}
{\max_{j\in\mathcal{J}} |\widetilde{e}_j|}<
\frac{1+\Delta}{(1-\Delta)^2}\cdot
\sqrt{\frac{\sigma_x^2+\sigma_z^2}{\sigma_x^2}}
\right|{\bf x}\in A_\epsilon\right)=1
\label{eq:Remove_Delta}.
\end{equation}
Equation~\eqref{eq:Remove_Delta} holds for any~$\Delta>0$. We note that~$\sqrt{\frac{\sigma_x^2+\sigma_z^2}{\sigma_x^2}}>1$, and thus~$\frac{1-\Delta}{(1+\Delta)^2}\cdot
\sqrt{\frac{\sigma_x^2+\sigma_z^2}{\sigma_x^2}}>1$ for sufficiently small~$\Delta$. Therefore,
\begin{eqnarray}
&&\lim_{N\to\infty}\Pr\left(\left.\frac{\max_{i\in\mathcal{I}} |e_i|}
{\max_{j\in\mathcal{J}} |\widetilde{e}_j|}>1\right|{\bf x}\in A_\epsilon\right)\nonumber\\
&=&\lim_{N\to\infty}\Pr\left(\left.\frac{\max_{i\in\mathcal{I}} |x_i-\widehat{x}_{\text{W,BG},i}|}
{\max_{j\in\mathcal{J}} |x_j-\widehat{x}_{\text{W,BG},j}|}>1\right|{\bf x}\in A_\epsilon\right)\nonumber\\
&=&1,
\label{eq:err_ij}
\end{eqnarray}
and
\begin{equation}
\lim_{N\to\infty}\Pr\left(\left.\frac{\max_{i\in\mathcal{I}} |x_i-\widehat{x}_{\text{W,BG},i}|}
{\max_{j\in\mathcal{J}} |x_j-\widehat{x}_{\text{W,BG},j}|}\le1\right|{\bf x}\in A_\epsilon\right)
=0.\nonumber
\end{equation}

{\bf Mean~$\ell_\infty$-norm error: }The road map for the remainder of the proof is to first show that when~${\bf x}\in A_\epsilon$ the Wiener filter is asymptotically optimal for expected~$\ell_\infty$-norm error, and then show that $\Pr({\bf x}\in A_\epsilon)$ is arbitrarily close to 1.

In order to utilize equations~\eqref{eq:Bounded_exp_1} and~\eqref{eq:Bounded_exp_2}, we normalize the quantities in the following derivations by~$\sqrt{\ln(N)}$ so that every term is bounded.
\begin{eqnarray}
&&\lim_{N\to\infty}\frac{E\left[\left.\|\mathbf{x}-\widehat{\mathbf{x}}_\text{W,BG}\|_\infty
\right|{\bf x}\in A_{\epsilon}\right]}{\sqrt{\ln(N)}}\nonumber\\
&=&\lim_{N\to\infty} 
E\left[\left.\frac{\max_{i\in\mathcal{I}}|{ x_i-\widehat{x}_{\text{W,BG},i}}|}
{\sqrt{\ln(N)}}
\right|{\bf x}\in  A_{\epsilon},\frac{\max_{i\in\mathcal{I}} |x_i-\widehat{x}_{\text{W,BG},i}|}
{\max_{j\in\mathcal{J}} |x_j-\widehat{x}_{\text{W,BG},j}|}>1\right]\cdot
\Pr\left(\left.\frac{\max_{i\in\mathcal{I}} |x_i-\widehat{x}_{\text{W,BG},i}|}
{\max_{j\in\mathcal{J}} |x_j-\widehat{x}_{\text{W,BG},j}|}>1\right|{\bf x}\in A_\epsilon\right)\nonumber\\
&+&\lim_{N\to\infty}
E\left[\left.\frac{\max_{j\in\mathcal{J}}|{ x_j-\widehat{x}_{\text{W,BG},j}}|}
{\sqrt{\ln(N)}}
\right|{\bf x}\in  A_{\epsilon},\frac{\max_{i\in\mathcal{I}} |x_i-\widehat{x}_{\text{W,BG},i}|}
{\max_{j\in\mathcal{J}} |x_j-\widehat{x}_{\text{W,BG},j}|}\le1\right]\cdot
\Pr\left(\left.\frac{\max_{i\in\mathcal{I}} |x_i-\widehat{x}_{\text{W,BG},i}|}
{\max_{j\in\mathcal{J}} |x_j-\widehat{x}_{\text{W,BG},j}|}\le1\right|{\bf x}\in A_\epsilon\right).\nonumber\\
\label{eq:remove_times_zero}
\end{eqnarray}
Let us now verify that the second term in~\eqref{eq:remove_times_zero} equals 0.
In fact, the following derivations hold from~\eqref{eq:maxE02} and~\eqref{eq:err_ij},
\begin{eqnarray}
1&=&\lim_{N\rightarrow\infty}\Pr\left(\left.\left|\frac{\max_{j\in\mathcal{J}}\left|x_j-\widehat{x}_{\text{W,GB},j}\right|}{\sqrt{2\cdot\frac{\sigma_x^4\sigma_z^2}{(\sigma_x^2+\sigma_z^2)^2}\cdot\ln (|\mathcal{J}|)}}-1\right|<\Delta\right|{\bf x}\in A_\epsilon\right)\nonumber\\
&=&\lim_{N\rightarrow\infty}\Pr\left(\left.\left|\frac{\max_{j\in\mathcal{J}}\left|x_j-\widehat{x}_{\text{W,GB},j}\right|}{\sqrt{2\cdot\frac{\sigma_x^4\sigma_z^2}{(\sigma_x^2+\sigma_z^2)^2}\cdot\ln (|\mathcal{J}|)}}-1\right|<\Delta\right|{\bf x}\in A_\epsilon,
\frac{\max_{i\in\mathcal{I}} |x_i-\widehat{x}_{\text{W,BG},i}|}
{\max_{j\in\mathcal{J}} |x_j-\widehat{x}_{\text{W,BG},j}|}>1
\right).\nonumber
\end{eqnarray}
Therefore,
\begin{equation}
\lim_{N\rightarrow\infty}E\left[\left.\frac{\max_{j\in\mathcal{J}}\left|x_j-\widehat{x}_{\text{W,GB},j}\right|}{\sqrt{2\cdot\frac{\sigma_x^4\sigma_z^2}{(\sigma_x^2+\sigma_z^2)^2}\cdot\ln (|\mathcal{J}|)}}\right|{\bf x}\in A_\epsilon,
\frac{\max_{i\in\mathcal{I}} |x_i-\widehat{x}_{\text{W,BG},i}|}
{\max_{j\in\mathcal{J}} |x_j-\widehat{x}_{\text{W,BG},j}|}>1
\right]=1,\nonumber
\end{equation}
which yields (following similar derivations of~\eqref{eq:Bounded_2} and~\eqref{eq:Bounded_exp_2})
\begin{equation}
\lim_{N\rightarrow\infty}E\left[\left.\frac{\max_{j\in\mathcal{J}}\left|x_j-\widehat{x}_{\text{W,GB},j}\right|}{\sqrt{\ln(N)}}\right|{\bf x}\in A_\epsilon,
\frac{\max_{i\in\mathcal{I}} |x_i-\widehat{x}_{\text{W,BG},i}|}
{\max_{j\in\mathcal{J}} |x_j-\widehat{x}_{\text{W,BG},j}|}>1
\right]=\sqrt{2\cdot\frac{\sigma_x^4\sigma_z^2}{(\sigma_x^2+\sigma_z^2)^2}}.
\nonumber
\end{equation}
Therefore, 
the second term in~\eqref{eq:remove_times_zero} equals $\sqrt{2\cdot\frac{\sigma_x^4\sigma_z^2}{(\sigma_x^2+\sigma_z^2)^2}}\times 0 = 0$, and equation~\eqref{eq:remove_times_zero} becomes
\begin{eqnarray}
&&\lim_{N\to\infty}\frac{E\left[\left.\|\mathbf{x}-\widehat{\mathbf{x}}_\text{W,BG}\|_\infty
\right|{\bf x}\in A_{\epsilon}\right]}{\sqrt{\ln(N)}}\nonumber\\
&=&\lim_{N\to\infty} 
E\left[\left.\frac{\max_{i\in\mathcal{I}}|{ x_i-\widehat{x}_{\text{W,BG},i}}|}{\sqrt{\ln(N)}}\right|{\bf x}\in  A_{\epsilon},\frac{\max_{i\in\mathcal{I}} |x_i-\widehat{x}_{\text{W,BG},i}|}
{\max_{j\in\mathcal{J}} |x_j-\widehat{x}_{\text{W,BG},j}|}>1\right]\nonumber\\
&=&\lim_{N\to\infty}E\left[\left.\frac{\max_{i\in\mathcal{I}}|{ x_i-\widehat{x}_{\text{W,BG},i}}|}
{\sqrt{\ln(N)}}
\right|{\bf x}\in A_{\epsilon}\right]\nonumber\\
&-&\lim_{N\to\infty}E\left[\left.\frac{\max_{i\in\mathcal{I}}|{ x_i-\widehat{x}_{\text{W,BG},i}}|}
{\sqrt{\ln(N)}}
\right|{\bf x}\in A_{\epsilon},
\frac{\max_{i\in\mathcal{I}} |x_i-\widehat{x}_{\text{W,BG},i}|}
{\max_{j\in\mathcal{J}} |x_j-\widehat{x}_{\text{W,BG},j}|}\le1\right]\times
\Pr\left(\left.\frac{\max_{i\in\mathcal{I}} |x_i-\widehat{x}_{\text{W,BG},i}|}
{\max_{j\in\mathcal{J}} |x_j-\widehat{x}_{\text{W,BG},j}|}\le1\right|{\bf x}\in A_\epsilon\right)\nonumber\\
&=&\lim_{N\to\infty}E\left[\left.\frac{\max_{i\in\mathcal{I}}|{ x_i-\widehat{x}_{\text{W,BG},i}}|}
{\sqrt{\ln(N)}}
\right|{\bf x}\in A_{\epsilon}\right].
\label{eq:wiener_nonzero}
\end{eqnarray}
Equation~\eqref{eq:wiener_nonzero} shows that the maximum absolute error of the Wiener filter relates to the Gaussian-distributed components of~${\bf x}$. 

{\bf Optimality of the Wiener filter: }It has been shown by Sherman~\cite{Sherman1955,Sherman1958} that, for parallel Gaussian channels with an i.i.d. Gaussian input~${\bf x}$, if an error metric function~$d({\bf x},\widehat{\bf x})$ relating~${\bf x}$ and its estimate~${\bf\widehat{x}}$ is convex, then the Wiener filter is optimal for that error metric. The~$\ell_\infty$-norm is convex, and therefore,
for any estimator $\mathbf{\widehat{x}}$,
\begin{eqnarray}
&&E\left[\left.\|{\bf x-\widehat{x}}\|_\infty\right|{\bf x}\in  A_{\epsilon}\right]\nonumber\\
&=&E\left[\left.\max_{i\in\mathcal{I\cup J}}|x_i-\widehat{x}_i|\right|{\bf x}\in  A_{\epsilon}\right]\nonumber\\
&\ge&E\left[\left.\max_{i\in\mathcal{I}}|x_i-\widehat{x}_i|\right|{\bf x}\in  A_{\epsilon}\right]\nonumber\\
&\ge&E\left[\left.\max_{i\in\mathcal{I}}|{ x_i-\widehat{x}_{\text{W,BG},i}}|\right|{\bf x}\in  A_{\epsilon}\right]\label{eq:opt_VS_wiener}.
\end{eqnarray}
The inequality~\eqref{eq:opt_VS_wiener} holds, because the set~$\{x_i:i\in\mathcal{I}\}$ only contains the i.i.d. Gaussian components of~${\bf x}$, and the Wiener filter is optimal for~$\ell_\infty$-norm error when the input signal is i.i.d. Gaussian.
The inequality~\eqref{eq:opt_VS_wiener} holds for any signal length N, and thus it holds when~$N\to\infty$ 
and we divide both sides by~$\sqrt{\ln(N)}$,
\begin{eqnarray}
0&\le&\lim_{N\to\infty}\left(\frac{E\left[\left.\|{\bf x-\widehat{x}}\|_\infty\right|{\bf x}\in A_{\epsilon}\right]}{\sqrt{\ln(N)}}
-\frac{E\left[\left.\max_{i\in\mathcal{I}}|{ x_i-\widehat{x}_{\text{W,BG},i}}|\right|{\bf x}\in A_{\epsilon}\right]}{\sqrt{\ln(N)}}\right)\nonumber\\
&=&\lim_{N\to\infty}\left(\frac{E\left[\left.
\|{\bf x-\widehat{x}}\|_\infty
\right|{\bf x}\in A_{\epsilon}\right]}
{\sqrt{\ln(N)}}-
\frac{E\left[\left.\|\mathbf{x}-\widehat{\mathbf{x}}_\text{W,BG}\|_\infty
\right|{\bf x}\in A_{\epsilon}\right]}{\sqrt{\ln(N)}}\right),
\label{eq:opt_VS_wiener02}
\end{eqnarray}
where the last step in~\eqref{eq:opt_VS_wiener02} is justified by the derivation of~\eqref{eq:wiener_nonzero}.
Equation~\eqref{eq:opt_VS_wiener02} also holds for~$\widehat{\bf x}=\widehat{\bf x}_{\ell_\infty}$,
\begin{equation}
\lim_{N\to\infty}\left(\frac{E\left[\left.
\|{\bf x-\widehat{x}_{\ell_\infty}}\|_\infty
\right|{\bf x}\in A_{\epsilon}\right]}{\sqrt{\ln(N)}}
-
\frac{E\left[\left.\|\mathbf{x}-\widehat{\mathbf{x}}_\text{W,BG}\|_\infty
\right|{\bf x}\in A_{\epsilon}\right]}{\sqrt{\ln(N)}}\right)\ge0.
\label{eq:W_opti_Aeps}
\end{equation}

{\bf Typical set:} Let us now evaluate~$\Pr({\bf x}\in A_\epsilon)$. 
The set~$A_\epsilon$ only considers whether the components in ${\bf x}$ are Gaussian or zero, and so 
we introduce a binary vector~${\bf \widetilde{x}}\in\mathbb{R}^N$, where~$\widetilde{x}_i=\mathbf{1}_{\left\{x_i\sim\mathcal{N}(\mu_x\sigma_x^2)\right\}}$ and~$\mathbf{1}_{\{\cdot\}}$ is the indicator function. That is,~$\widetilde{x}_i=1$ if~$x_i$ is Gaussian, and else~$\widetilde{x}_i=0$.
The sequence~${\bf\widetilde{x}}\triangleq\{\widetilde{x}_1,\widetilde{x}_2,\ldots,\widetilde{x}_N\}$ is called a {\em typical sequence} (\cite{Cover06}, page 59), if it satisfies
\begin{equation}
2^{-N(H(\widetilde{\bf X})+\delta)}\le\Pr(\widetilde{x}_1,\widetilde{x}_2,\ldots,\widetilde{x}_N)\le2^{-N(H(\widetilde{\bf X})-\delta)},
\label{eq:TypSeq}
\end{equation}
for some~$\delta>0$, where~$H(\widetilde{\bf X})$ denotes the binary entropy~\cite{Cover06} of the sequence~$\{\widetilde{x}_1,\widetilde{x}_2,\ldots,\widetilde{x}_N\}$.
The set~$A_\epsilon$ is then called a {\em typical set}~\cite{Cover06}, and
\begin{equation}
\Pr({\bf x}\in A_\epsilon)>1-\delta.\label{eq:TypSet}
\end{equation}
We highlight that the inequalities~\eqref{eq:TypSeq} and~\eqref{eq:TypSet} both hold when~$\delta\to0^+$ as a function of~$N$.

In our problem setting where~$\Pr(\widetilde{x}_i=1)=\Pr(x_i\sim\mathcal{N}(\mu_x,\sigma_x^2))=s$, the entropy of the sequence~$\{\widetilde{x}_1,\widetilde{x}_2,\ldots,\widetilde{x}_N\}$ is
\begin{equation}
H(\widetilde{\bf X})=-s\log_2(s)-(1-s)\log_2(1-s),\label{eq:HX}
\end{equation}
and the probability of the sequence~$\{\widetilde{x}_1,\widetilde{x}_2,\ldots,\widetilde{x}_N\}$ is
\begin{equation}
\Pr(\widetilde{x}_1,\widetilde{x}_2,\ldots,\widetilde{x}_N)=s^{|\mathcal{I}|}\cdot (1-s)^{|\mathcal{J}|}.
\label{eq:prob_X}
\end{equation}
Plugging~\eqref{eq:IJ_ineq},~\eqref{eq:HX}, and~\eqref{eq:prob_X} into~\eqref{eq:TypSeq}, the value of~$\delta$ can be computed,
\begin{equation}
\delta=\epsilon\left|\log_2\left(\frac{s}{1-s}\right)\right|,
\label{eq:delta_eps}
\end{equation}
for~$0<s<1$ and~$s\neq0.5$. That is,
\begin{equation}
\Pr({\bf x}\in A_\epsilon)>1-\delta=1-\epsilon\left|\log_2\left(\frac{s}{1-s}\right)\right|.\label{eq:PrA}
\end{equation}

Finally, we compare~$E\left[\|\mathbf{x}-\widehat{\mathbf{x}}_\text{WB,G}\|_\infty\right]$ with~$E\left[\|\mathbf{x}-\widehat{\mathbf{x}}_{\ell_\infty}\|_\infty\right]$, where~$\widehat{\bf x}_{\ell_\infty}$ satisfies~\eqref{eq:ell_infty_estimator}, i.e., the estimate $\widehat{\bf x}_{\ell_\infty}$ is optimal for minimizing the mean~$\ell_\infty$-norm error of estimation. By definition,
\begin{equation}
\lim_{N\to\infty}
\left(
\frac{E\left[\|{\bf x}-\widehat{\bf x}_{\ell_\infty}\|_\infty\right]}{\sqrt{\ln(N)}}-
\frac{E\left[\|{\bf x}-\widehat{\bf x}_\text{W,BG}\|_\infty\right]}{\sqrt{\ln(N)}}
\right)
\le0,\nonumber
\end{equation}
but we already proved in~\eqref{eq:W_opti_Aeps} that
\begin{equation}
\lim_{N\to\infty}
\left(
\frac{E\left[\left.\|{\bf x}-\widehat{\bf x}_{\ell_\infty}\|_\infty\right|{\bf x}\in A_\epsilon\right]}{\sqrt{\ln(N)}}-
\frac{E\left[\left.\|{\bf x}-\widehat{\bf x}_\text{W,BG}\|_\infty\right|{\bf x}\in A_\epsilon\right]}{\sqrt{\ln(N)}}
\right)
\ge0,\nonumber
\end{equation}
and thus
\begin{equation}
\lim_{N\to\infty}
\left(
\frac{E\left[\left.\|{\bf x}-\widehat{\bf x}_{\ell_\infty}\|_\infty\right|{\bf x}\notin A_\epsilon\right]}{\sqrt{\ln(N)}}-
\frac{E\left[\left.\|{\bf x}-\widehat{\bf x}_\text{W,BG}\|_\infty\right|{\bf x}\notin A_\epsilon\right]}{\sqrt{\ln(N)}}
\right)
\le0.
\label{eq:neq_A_c}
\end{equation}
We know that~$\Pr\left({\bf x}\notin A_\epsilon\right)<\delta$ from~\eqref{eq:PrA}. 
To complete the proof, it suffices to show that, when~${\bf x}\notin A_\epsilon$, the subtraction~\eqref{eq:neq_A_c} is bounded.
When~${\bf x}\notin A_\epsilon$, there are 3 cases for the possible values of~$|\mathcal{I}|$ and~$|\mathcal{J}|$:
\begin{itemize}
\item Case 1: $|\mathcal{I}|,|\mathcal{J}|\to\infty$, but~\eqref{eq:ratio} may not hold.
\item Case 2: $|\mathcal{J}|\to\infty$ but $|\mathcal{I}|\not\to\infty$.
\item Case 3: $|\mathcal{I}|\to\infty$ but $|\mathcal{J}|\not\to\infty$.
\end{itemize}

We observe that equations~\eqref{eq:Bounded_exp_1} and~\eqref{eq:Bounded_exp_2} are derived from~\eqref{eq:Bounded_1},~\eqref{eq:Bounded_2}, and~\eqref{eq:ratio}. In Case 1, similar equalities to~\eqref{eq:Bounded_1} and~\eqref{eq:Bounded_2} hold,
\begin{equation}
\lim_{N\to\infty} 
E\left[\left.\frac{\max_{i\in\mathcal{I}}|e_i|}{\sqrt{
\ln(N)}}\right|\text{Case 1 of }{\bf x}\notin A_\epsilon\right]=
\lim_{N\to\infty} \sqrt{2
\cdot\frac{\sigma_x^2\sigma_z^2}{\sigma_x^2+\sigma_z^2}\cdot
\frac{\ln(|\mathcal{I}|)}{\ln(N)}}
\le
\sqrt{2
\cdot\frac{\sigma_x^2\sigma_z^2}{\sigma_x^2+\sigma_z^2}}\nonumber
\end{equation}
and
\begin{equation}
\lim_{N\to\infty} 
E\left[\left.\frac{\max_{j\in\mathcal{J}}|\widetilde{e}_j|}{\sqrt{
\ln(N)}}\right|\text{Case 1 of }{\bf x}\notin A_\epsilon\right]=
\lim_{N\to\infty} \sqrt{2
\cdot\frac{\sigma_x^4\sigma_z^2}{(\sigma_x^2+\sigma_z^2)^2}\cdot
\frac{\ln(|\mathcal{J}|)}{\ln(N)}}
\le
\sqrt{2
\cdot\frac{\sigma_x^4\sigma_z^2}{(\sigma_x^2+\sigma_z^2)^2}}.\nonumber
\end{equation}
Therefore, the value of~$
\lim_{N\to\infty} 
E\left[\left.\frac{\|{\bf x}-\widehat{\bf x}_\text{W,BG}\|_\infty}{\sqrt{
\ln(N)}}\right|\text{Case 1 of }{\bf x}\notin A_\epsilon\right]
$ is bounded.

In Case 2, it is obvious that~$\lim_{N\to\infty} 
E\left[\left.\frac{\max_{i\in\mathcal{I}}|e_i|}{\sqrt{
\ln(N)}}\right|\text{Case 2 of }{\bf x}\notin A_\epsilon\right]$ is bounded because~$|\mathcal{I}|\not\to\infty$, while $\lim_{N\to\infty} 
E\left[\left.\frac{\max_{j\in\mathcal{J}}|\widetilde{e}_j|}{\sqrt{
\ln(N)}}\right|\text{Case 2 of }{\bf x}\notin A_\epsilon\right]$ is bounded because~$|\mathcal{J}|\le N$, and
\begin{equation}
\lim_{N\to\infty} 
E\left[\left.\frac{\max_{j\in\mathcal{J}}|\widetilde{e}_j|}{\sqrt{
\ln(N)}}\right|\text{Case 2 of }{\bf x}\notin A_\epsilon\right]\le
\sqrt{2
\cdot\frac{\sigma_x^4\sigma_z^2}{(\sigma_x^2+\sigma_z^2)^2}}.\nonumber
\end{equation} 
The analysis for Case 3 is similar to that of Case 2.

Therefore, we have shown that~$
\lim_{N\to\infty} 
E\left[\left.\frac{\|{\bf x}-\widehat{\bf x}_\text{W,BG}\|_\infty}{\sqrt{
\ln(N)}}\right|{\bf x}\notin A_\epsilon\right]
$ is bounded.

By~\eqref{eq:neq_A_c}, $\lim_{N\to\infty}\frac{E\left[\left.\|\mathbf{x}-\widehat{\mathbf{x}}_{\ell_\infty}\|_\infty
\right|{\bf x}\notin A_{\epsilon}\right]}{\sqrt{\ln(N)}}$ is bounded above by $
\lim_{N\to\infty} 
E\left[\left.\frac{\|{\bf x}-\widehat{\bf x}_\text{W,BG}\|_\infty}{\sqrt{
\ln(N)}}\right|{\bf x}\notin A_\epsilon\right]
$.
Hence,
\begin{equation}
\lim_{N\to\infty}\left(
\frac{E\left[\left.\|\mathbf{x}-\widehat{\mathbf{x}}_\text{W,BG}\|_\infty
\right|{\bf x}\notin A_{\epsilon}\right]}{\sqrt{\ln(N)}}-
\frac{E\left[\left.\|\mathbf{x}-\widehat{\mathbf{x}}_{\ell_\infty}\|_\infty
\right|{\bf x}\notin A_{\epsilon}\right]}{\sqrt{\ln(N)}}
\right)
=c\nonumber
\end{equation}
is bounded, where~$c>0$ is a constant. 

Therefore,
\begin{eqnarray}
&&\lim_{N\to\infty}\frac{E\left[\|\mathbf{x}-\widehat{\mathbf{x}}_\text{W,BG}\|_\infty
\right]}
{E\left[\|\mathbf{x}-\widehat{\mathbf{x}}_{\ell_\infty}\|_\infty
\right]}\nonumber\\
&=&
\lim_{N\to\infty}
\frac{\frac{E\left[\|\mathbf{x}-\widehat{\mathbf{x}}_\text{W,BG}\|_\infty
\right]}{\sqrt{\ln(N)}}}
{\frac{E\left[\|\mathbf{x}-\widehat{\mathbf{x}}_{\ell_\infty}\|_\infty
\right]}{\sqrt{\ln(N)}}}\nonumber\\
&=&
\lim_{N\to\infty}
\frac{\frac{E\left[\left.\|\mathbf{x}-\widehat{\mathbf{x}}_\text{W,BG}\|_\infty
\right|{\bf x}\in A_{\epsilon}\right]}{\sqrt{\ln(N)}}\cdot\Pr\left({\bf x}\in A_\epsilon\right)+
\frac{E\left[\left.\|\mathbf{x}-\widehat{\mathbf{x}}_\text{W,BG}\|_\infty
\right|{\bf x}\notin A_{\epsilon}\right]}{\sqrt{\ln(N)}}\cdot\Pr\left({\bf x}\notin A_\epsilon\right)}
{\frac{E\left[\left.\|\mathbf{x}-\widehat{\mathbf{x}}_{\ell_\infty}\|_\infty
\right|{\bf x}\in A_{\epsilon}\right]}{\sqrt{\ln(N)}}\cdot\Pr\left({\bf x}\in A_\epsilon\right)+
\frac{E\left[\left.\|\mathbf{x}-\widehat{\mathbf{x}}_{\ell_\infty}\|_\infty
\right|{\bf x}\notin A_{\epsilon}\right]}{\sqrt{\ln(N)}}\cdot\Pr\left({\bf x}\notin A_\epsilon\right)}\nonumber
\end{eqnarray}
\begin{eqnarray}
&=&
\lim_{N\to\infty}
\frac{\frac{E\left[\left.\|\mathbf{x}-\widehat{\mathbf{x}}_\text{W,BG}\|_\infty
\right|{\bf x}\in A_{\epsilon}\right]}{\sqrt{\ln(N)}}\cdot\Pr\left({\bf x}\in A_\epsilon\right)+
\frac{E\left[\left.\|\mathbf{x}-\widehat{\mathbf{x}}_{\ell_\infty}\|_\infty
\right|{\bf x}\notin A_{\epsilon}\right]}{\sqrt{\ln(N)}}\cdot
\Pr\left({\bf x}\notin A_\epsilon\right)
}
{\frac{E\left[\left.\|\mathbf{x}-\widehat{\mathbf{x}}_{\ell_\infty}\|_\infty
\right|{\bf x}\in A_{\epsilon}\right]}{\sqrt{\ln(N)}}\cdot\Pr\left({\bf x}\in A_\epsilon\right)+
\frac{E\left[\left.\|\mathbf{x}-\widehat{\mathbf{x}}_{\ell_\infty}\|_\infty
\right|{\bf x}\notin A_{\epsilon}\right]}{\sqrt{\ln(N)}}\cdot
\Pr\left({\bf x}\notin A_\epsilon\right)}
\nonumber\\
&&+
\lim_{N\to\infty}
\frac{
\frac{E\left[\left.\|\mathbf{x}-\widehat{\mathbf{x}}_\text{W,BG}\|_\infty
\right|{\bf x}\notin A_{\epsilon}\right]}{\sqrt{\ln(N)}}\cdot
\Pr\left({\bf x}\notin A_\epsilon\right)
-\frac{E\left[\left.\|\mathbf{x}-\widehat{\mathbf{x}}_{\ell_\infty}\|_\infty
\right|{\bf x}\notin A_{\epsilon}\right]}{\sqrt{\ln(N)}}\cdot
\Pr\left({\bf x}\notin A_\epsilon\right)
}
{\frac{E\left[\left.\|\mathbf{x}-\widehat{\mathbf{x}}_{\ell_\infty}\|_\infty
\right|{\bf x}\in A_{\epsilon}\right]}{\sqrt{\ln(N)}}\cdot
\Pr\left({\bf x}\in A_\epsilon\right)+
\frac{E\left[\left.\|\mathbf{x}-\widehat{\mathbf{x}}_{\ell_\infty}\|_\infty
\right|{\bf x}\notin A_{\epsilon}\right]}{\sqrt{\ln(N)}}\cdot
\Pr\left({\bf x}\notin A_\epsilon\right)}\nonumber\\
&\le&1+
\lim_{N\to\infty}
\frac{\left(
\frac{E\left[\left.\|\mathbf{x}-\widehat{\mathbf{x}}_\text{W,BG}\|_\infty
\right|{\bf x}\notin A_{\epsilon}\right]}{\sqrt{\ln(N)}}
-\frac{E\left[\left.\|\mathbf{x}-\widehat{\mathbf{x}}_{\ell_\infty}\|_\infty
\right|{\bf x}\notin A_{\epsilon}\right]}{\sqrt{\ln(N)}}\right)
\cdot\Pr\left({\bf x}\notin A_\epsilon\right)
}
{\frac{E\left[\|\mathbf{x}-\widehat{\mathbf{x}}_{\ell_\infty}\|_\infty
\right]}{\sqrt{\ln(N)}}}\nonumber\\
&<&1+
\lim_{N\to\infty}
\frac{c\cdot\delta
}
{\frac{E\left[\|\mathbf{x}-\widehat{\mathbf{x}}_{\ell_\infty}\|_\infty
\right]}{\sqrt{\ln(N)}}}\label{eq:close_to_final}.
\end{eqnarray}
{In~\eqref{eq:close_to_final}, 
the value of~$\lim_{N\to\infty}\frac{E\left[\|\mathbf{x}-\widehat{\mathbf{x}}_{\ell_\infty}\|_\infty
\right]}{\sqrt{\ln(N)}}$ is bounded below because of~\eqref{eq:W_opti_Aeps},
\begin{eqnarray}
&&\lim_{N\to\infty}\frac{E\left[\|\mathbf{x}-\widehat{\mathbf{x}}_{\ell_\infty}\|_\infty
\right]}{\sqrt{\ln(N)}}\nonumber\\
&=&\lim_{N\to\infty}\frac{E\left[\left.\|\mathbf{x}-\widehat{\mathbf{x}}_{\ell_\infty}\|_\infty\right|{\bf x}\in A_\epsilon
\right]}{\sqrt{\ln(N)}}\cdot\Pr\left({\bf x}\in A_\epsilon\right)
+\lim_{N\to\infty}\frac{E\left[\left.\|\mathbf{x}-\widehat{\mathbf{x}}_{\ell_\infty}\|_\infty\right|{\bf x}\notin A_\epsilon
\right]}{\sqrt{\ln(N)}}\cdot\Pr\left({\bf x}\notin A_\epsilon\right)\nonumber\\
&\ge&
\lim_{N\to\infty}\frac{E\left[\left.\|\mathbf{x}-\widehat{\mathbf{x}}_\text{W,BG}\|_\infty\right|{\bf x}\in A_\epsilon
\right]}{\sqrt{\ln(N)}}\cdot\Pr\left({\bf x}\in A_\epsilon\right)\nonumber\\
&>&\sqrt{2
\cdot\frac{\sigma_x^2\sigma_z^2}{\sigma_x^2+\sigma_z^2}}\cdot(1-\delta).\nonumber
\end{eqnarray}
On the other hand, whether the value of~$\frac{E\left[\|\mathbf{x}-\widehat{\mathbf{x}}_{\ell_\infty}\|_\infty
\right]}{\sqrt{\ln(N)}}$ is bounded above or not, the second term in~\eqref{eq:close_to_final} is always arbitrarily small because~$\delta$ is arbitrarily small, and thus~\eqref{eq:close_to_final} is equivalent to
\begin{equation}
\lim_{N\to\infty}\frac{E\left[\|\mathbf{x}-\widehat{\mathbf{x}}_\text{W,BG}\|_\infty
\right]}
{E\left[\|\mathbf{x}-\widehat{\mathbf{x}}_{\ell_\infty}\|_\infty
\right]}
<1+\delta,\nonumber
\end{equation}
where~$\delta\to0^+$ as a function of~$N$.
Finally, because~$\widehat{\bf x}_\infty$ is the optimal estimator for~$\ell_\infty$-norm error,
\begin{equation}
\lim_{N\to\infty}\frac{E\left[\|\mathbf{x}-\widehat{\mathbf{x}}_\text{W,BG}\|_\infty
\right]}
{E\left[\|\mathbf{x}-\widehat{\mathbf{x}}_{\ell_\infty}\|_\infty
\right]}
\ge1.\nonumber
\end{equation}
Therefore,
\begin{equation}
\lim_{N\to\infty}\frac{E\left[\|\mathbf{x}-\widehat{\mathbf{x}}_\text{W,BG}\|_\infty
\right]}
{E\left[\|\mathbf{x}-\widehat{\mathbf{x}}_{\ell_\infty}\|_\infty
\right]}
=1,\nonumber
\end{equation}
which completes the proof.

\subsection{Proof of Theorem~\ref{coro:01}}
\label{appen:coro_01}

The road map of the proof of Theorem~\ref{coro:01} is the same as that of Theorem~\ref{thm:01}. 

{\bf $K$ error patterns: }The input signal of the parallel Gaussian channels~\eqref{eq:scalar} is generated by an i.i.d. Gaussian mixture source~\eqref{eq:MGauss}, and suppose without loss of generality that~$\sigma_1^2=\max_{k\in\{1,2,\ldots,K\}}\sigma_k^2$. The Wiener filter is 
$\mathbf{\widehat{x}}_\text{W,GM}=\frac{\sigma_1^2}{\sigma_1^2+\sigma_z^2}\cdot (\mathbf{r}-\mu_1)+\mu_1=\frac{\sigma_1^2{\bf r}+\sigma_z^2\mu_1}{\sigma_1^2+\sigma_z^2}$. 
Let $\mathcal{I}_k$ denote the index set where $x_i\sim\mathcal{N}(\mu_k,\sigma_k^2)$,. Then we define~$K$ types of error patterns: for~$k\in\{1,2,\ldots,K\}$, the~$k$-th error pattern is
\begin{equation}
e^{(k)}_{i}\triangleq\widehat{x}_{\text{W,GM},i}-x_i=\frac{\sigma_1^2 r_i+\sigma_z^2\mu_1}{\sigma_1^2+\sigma_z^2}-x_i
\sim\mathcal{N}\left(\frac{\sigma_z^2}{\sigma_1^2+\sigma_z^2}\mu_1-
\frac{\sigma_z^2}{\sigma_1^2+\sigma_z^2}\mu_k,
\frac{\sigma_z^4}{(\sigma_1^2+\sigma_z^2)^2}\sigma_k^2+
\frac{\sigma_1^4}{(\sigma_1^2+\sigma_z^2)^2}\sigma_z^2\right)
\nonumber,
\end{equation}
where
\begin{equation}
i\in\mathcal{I}_k\triangleq\{i:x_i\sim\mathcal{N}(\mu_k,\sigma_k^2)\}.
\nonumber
\end{equation}
Because the variances~$\sigma_z^2,\sigma_1^2,\sigma_2^2,\ldots,\sigma_K^2>0$ are constants, and~$\sigma_1^2=\max_{k\in\{1,2,\ldots,K\}}\sigma_k^2$,
\begin{equation}
\frac{\sigma_z^4}{(\sigma_1^2+\sigma_z^2)^2}\sigma_1^2+
\frac{\sigma_1^4}{(\sigma_1^2+\sigma_z^2)^2}\sigma_z^2
=\max_{k\in\{1,2,\ldots,K\}}\left(
\frac{\sigma_z^4}{(\sigma_1^2+\sigma_z^2)^2}\sigma_k^2+
\frac{\sigma_1^4}{(\sigma_1^2+\sigma_z^2)^2}\sigma_z^2\right),\label{eq:max_ek}
\end{equation}
which shows that the first error pattern~$e_i^{(1)}$ has the greatest variance.

{\bf Maximum of error patterns: }Define the set~$A_\epsilon$ as
\begin{equation}
A_\epsilon\triangleq\left\{{\bf x}:
\left|\frac{|\mathcal{I}_1|}{N}-s_1\right|<\epsilon_1,
\left|\frac{|\mathcal{I}_2|}{N}-s_2\right|<\epsilon_2,\ldots,
\left|\frac{|\mathcal{I}_K|}{N}-s_K\right|<\epsilon_K\right\},\nonumber
\end{equation}
where~$\sum_{k=1}^K|\mathcal{I}_k|=N$, 
and~$\epsilon_k\to0^+$ as a function of~$N$ for~$k\in\{1,2,\ldots,K\}$.
Applying a similar derivation to that of~\eqref{eq:err_ij}, we obtain that
\begin{eqnarray}
&&\lim_{N\to\infty}\Pr\left(\left.\frac{\max_{i\in\mathcal{I}_1} |\widehat{x}_{\text{W,GM},i}-x_i|}{\max_{j\in\mathcal{I}_k} |\widehat{x}_{\text{W,GM},j}-x_j|}>
\frac{1-\Delta}{(1+\Delta)^2}\cdot
\frac{\sqrt{\frac{\sigma_z^4}{(\sigma_1^2+\sigma_z^2)^2}\sigma_1^2+
\frac{\sigma_1^4}{(\sigma_1^2+\sigma_z^2)^2}\sigma_z^2}}
{\sqrt{\frac{\sigma_z^4}{(\sigma_1^2+\sigma_z^2)^2}\sigma_k^2+
\frac{\sigma_1^4}{(\sigma_1^2+\sigma_z^2)^2}\sigma_z^2}}\right|{\bf x}\in A_\epsilon\right)\label{eq:eq01}\\
&=&\lim_{N\to\infty}\Pr\left(\left.\frac{\max_{i\in\mathcal{I}_1} |\widehat{x}_{\text{W,GM},i}-x_i|}{\max_{j\in\mathcal{I}_k} |\widehat{x}_{\text{W,GM},j}-x_j|}\ge1\right|{\bf x}\in A_\epsilon\right)\label{eq:eq02}\\
&=&1,\nonumber
\end{eqnarray}
for any~$k\neq1$. Equation~\eqref{eq:eq02} is valid because~\eqref{eq:eq01} holds for any~$\Delta>0$, and~$\frac{\sqrt{\frac{\sigma_z^4}{(\sigma_1^2+\sigma_z^2)^2}\sigma_1^2+
\frac{\sigma_1^4}{(\sigma_1^2+\sigma_z^2)^2}\sigma_z^2}}
{\sqrt{\frac{\sigma_z^4}{(\sigma_1^2+\sigma_z^2)^2}\sigma_k^2+
\frac{\sigma_1^4}{(\sigma_1^2+\sigma_z^2)^2}\sigma_z^2}}\ge1$ is derived from~\eqref{eq:max_ek}.

Hence, 
\begin{equation}
\lim_{N\to\infty}E\left[
\left.\frac{\|\mathbf{x}-\widehat{\mathbf{x}}_\text{W,GM}\|_\infty}{\sqrt{\ln(N)}}\right|{\bf x}\in A_{\epsilon}\right]\lim_{N\to\infty}E\left[\left.\frac{
\max_{i\in\mathcal{I}_1}|{ x_i-\widehat{x}_{\text{W,GM},i}}|}{\sqrt{\ln(N)}}\right|{\bf x}\in A_{\epsilon}\right].
\label{eq:wiener_GM}
\end{equation}
Equation~\eqref{eq:wiener_GM} shows that the maximum absolute error of the Wiener filter relates to the Gaussian component that has the greatest variance.

{\bf Optimality of the Wiener filter: }Then applying similar derivations of equations~\eqref{eq:opt_VS_wiener02} and~\eqref{eq:W_opti_Aeps},
\begin{eqnarray}
\lim_{N\to\infty}\left(E\left[\left.\frac{
\|{\bf x-\widehat{x}_{\ell_\infty}}\|_\infty}{\sqrt{\ln(N)}}\right|{\bf x}\in A_{\epsilon}\right]
- E\left[\left.\frac{
\|\mathbf{x}-\widehat{\mathbf{x}}_\text{W,GM}\|_\infty}{\sqrt{\ln(N)}}\right|{\bf x}\in A_{\epsilon}\right]\right)\ge0.\nonumber
\end{eqnarray}

{\bf Typical set: }Similar to the derivation of~\eqref{eq:delta_eps}, we obtain the probability of~${\bf x}\in A_\epsilon$~(\cite{Cover06}, page 59),
\begin{equation}
\Pr({\bf x}\in A_\epsilon)>1-\delta,\nonumber
\end{equation}
where
\begin{equation}
\delta=\sum_{k=1}^K\epsilon_k\left|\log_2(s_k)\right|.\nonumber
\end{equation}
Finally,
\begin{equation}
\lim_{N\to\infty}\frac{E\left[\|\mathbf{x}-\widehat{\mathbf{x}}_\text{W,GM}\|_\infty
\right]}
{E\left[\|\mathbf{x}-\widehat{\mathbf{x}}_{\ell_\infty}\|_\infty
\right]}
<1+\delta,\nonumber
\end{equation}
where~$\delta\to0^+$, and thus
\begin{equation}
\lim_{N\to\infty}\frac{E\left[\|\mathbf{x}-\widehat{\mathbf{x}}_\text{W,GM}\|_\infty
\right]}
{E\left[\|\mathbf{x}-\widehat{\mathbf{x}}_{\ell_\infty}\|_\infty
\right]}
=1.\nonumber
\end{equation}

\subsection{Proof of Lemma~\ref{lemma}}
\label{append:lemma}

It has been shown~\cite{Gnedenko1943,Berman1962} that for an i.i.d. standard Gaussian sequence~$\widetilde{u}_i\sim\mathcal{N}(0,1)$, where~$i\in\{1,2,\ldots,N\}$, the maximum of the sequence,~$\max_i\widetilde{u}_i$, converges to~$\sqrt{2\ln(N)}$ in probability, i.e.,
\begin{equation}
\lim_{N\rightarrow\infty}\Pr\left(\left|\frac{\max_{1\le i\le N}\widetilde{u}_i}{\sqrt{2\cdot\ln (N)}}-1\right|<\Delta\right)=1,\nonumber
\end{equation}
for any~$\Delta>0$.
Therefore, for an i.i.d. non-standard Gaussian sequence~$u_i\sim\mathcal{N}(\mu, \sigma^2)$,~$\frac{u_i-\mu}{|\sigma|}\sim\mathcal{N}(0,1)$, and it follows that
\begin{equation}
\lim_{N\rightarrow\infty}\Pr\left(\left|\frac{\max_{1\le i\le N}(u_i-\mu)}{\sqrt{2\sigma^2\cdot\ln (N)}}-1\right|<\Delta\right)=1,\label{eq:arbG}
\end{equation}
for any~$\Delta>0$.
We observe that, for a given~$\mu$, the following probability equals 1 for sufficient large~$N$, and therefore,
\begin{equation}
\lim_{N\rightarrow\infty}\Pr\left(\left|\frac{-\mu}{\sqrt{2\sigma^2\cdot\ln (N)}}-0\right|<\Delta\right)=1,
\label{eq:arbG_mean}
\end{equation}
for any~$\Delta>0$. 
Combining~\eqref{eq:arbG} and~\eqref{eq:arbG_mean}, 
\begin{equation}
\lim_{N\rightarrow\infty}\Pr\left(\left|\frac{\max_{1\le i\le N}u_i}{\sqrt{2\sigma^2\cdot\ln (N)}}-1\right|<2\Delta\right)=1,\nonumber
\end{equation}
for any~$\Delta>0$, which owing to arbitrariness of~$\Delta$ yields
\begin{equation}
\lim_{N\rightarrow\infty}\Pr\left(\left|\frac{\max_{1\le i\le N}u_i}{\sqrt{2\sigma^2\cdot\ln (N)}}-1\right|<\Delta\right)=1.\label{eq:arbG_final}
\end{equation}
Equation~\eqref{eq:arbG_final} suggests that, for a sequence of i.i.d. Gaussian random variables, $u_i~\sim\mathcal{N}(\mu, \sigma^2)$, the maximum of the sequence is not affected by the value of~$\mu$.

On the other hand, the i.i.d. Gaussian sequence~$(-u_i)\sim\mathcal{N}(-\mu,\sigma^2)$ satisfies
\begin{equation}
\lim_{N\rightarrow\infty}\Pr\left(\left|\frac{\max_{1\le i\le N}(-u_i)}{\sqrt{2\sigma^2\cdot\ln (N)}}-1\right|<\Delta\right)=1.\nonumber
\end{equation}
Hence, 
\begin{eqnarray}
&&\lim_{N\rightarrow\infty}\Pr\left(\left|\frac{\max_{1\le i\le N}|u_i|}{\sqrt{2\sigma^2\cdot\ln (N)}}-1\right|<\Delta\right)\nonumber\\
&=&\lim_{N\to\infty}\Pr\left(\left|\frac{\max_{1\le i\le N}u_i}{\sqrt{2\sigma^2\cdot\ln (N)}}-1\right|<\Delta\text{ and }
\left|\frac{\max_{1\le i\le N}(-u_i)}{\sqrt{2\sigma^2\cdot\ln (N)}}-1\right|<\Delta\right)\nonumber\\
&=&\lim_{N\to\infty}\Pr\left(\left|\frac{\max_{1\le i\le N}u_i}{\sqrt{2\sigma^2\cdot\ln (N)}}-1\right|<\Delta\right)
-\lim_{N\to\infty}\Pr\left(\left|\frac{\max_{1\le i\le N}u_i}{\sqrt{2\sigma^2\cdot\ln (N)}}-1\right|<\Delta\text{ and }
\left|\frac{\max_{1\le i\le N}(-u_i)}{\sqrt{2\sigma^2\cdot\ln (N)}}-1\right|>\Delta\right)\nonumber\\
&=&\lim_{N\to\infty}\Pr\left(\left|\frac{\max_{1\le i\le N}u_i}{\sqrt{2\sigma^2\cdot\ln (N)}}-1\right|<\Delta\right)-0
\nonumber\\
&=&1,\nonumber
\end{eqnarray}
for any~$\Delta>0$.

\section{Conclusion}
\label{sec:concld}

This correspondence focused on estimating input signals in parallel Gaussian channels, where the signals were generated by i.i.d. Gaussian mixture sources, and the~$\ell_\infty$-norm error was used to quantify the performance. 
We proved that the Wiener filter~\eqref{eq:Wiener02}, a simple linear function that is applied to the Gaussian channel outputs, asymptotically minimizes the mean~$\ell_\infty$-norm error when the signal dimension~$N\to\infty$. 
Specifically, the multiplicative constant of the linear filter only relates to the greatest variance of the Gaussian mixture components and the variance of the Gaussian noise. Our results for parallel Gaussian channels can be extended to linear mixing systems, in settings where linear mixing systems can be decoupled to parallel Gaussian channels.

Our results are asymptotic, 
but one will notice from Section~\ref{appen:ThmProof}~\eqref{eq:J_over_I} that the asymptotic results hold only for astronomically large signal dimension~$N$, which may lead readers to wonder whether the Wiener filter performs well when the signal dimension is finite.
To answer this question, we performed numerical simulations for finite signal dimensions. The numerical results showed that the Wiener filter indeed reduces the~$\ell_\infty$-norm error to some extent. Specifically, the Wiener filter outperforms the relaxed belief propagation algorithm~\cite{Rangan2010CISS,Rangan2010} in linear mixing systems. However, our numerical results suggested that there exist better algorithms~\cite{Tan2014CISS} for~$\ell_\infty$-norm error than the Wiener filter in finite signal dimension settings. The development of optimal algorithms in the finite dimension setting is left for future work.

\section*{Acknowledgments}
We thank Nikhil Krishnan for useful discussions.
We also thank the reviewers for their comments, which greatly helped us improve this manuscript.

\ifCLASSOPTIONcaptionsoff
  \newpage
\fi



\bibliography{cites}

\end{document}